\documentstyle[preprint,aps,version2]{revtex}
\begin{document}
\draft
\preprint{OCIP/C 94-2}
\preprint{July 1994}
\preprint{Revised: August 1996}
\begin{title}
Probing the Heavy Quark Content of the Photon \\
Using $b$ Tagging
at High Energy  $e\gamma$ and $e^+e^-$ Colliders
\end{title}
\author{Michael A. Doncheski\thanks{Present address: Department 
of Physics, Pennsylvania State University, Mont Alto, PA 17237 USA}, 
Stephen Godfrey and \\
K. Andrew Peterson\thanks{Present address: Department of
Physics, Memorial University of Newfoundland, St. John's, NF, Canada,
A1B 3X7}}
\begin{instit}
Ottawa-Carleton Institute for Physics \\
Department of Physics, Carleton University, Ottawa CANADA, K1S 5B6
\end{instit}

\begin{abstract}
We suggest a method for probing the quark content of the photon 
using $b$ tagging  at high energy $e^+e^-$ and $e\gamma$ colliders.  
We find that heavy quark tagging provides a sensitive and effective 
probe of the quark content of the photon especially in the low 
$x$-region where the various models differ the most.  This process 
is complementary to others that have been studied in the literature 
and can contribute to more precise determinations of quark and gluon 
distributions inside the photon.
\end{abstract}

%\narrowtext
\section{INTRODUCTION}
\label{sec:intro}

There is a growing interest in the hadronic content of the photon
\cite{witten} as both a test of QCD and as a background to precision
measurements of electroweak parameters.  In fact, some 
photoproduction processes are dominated by the hadronic interactions 
of photons in certain regions of phase space \cite{eboli,photon}.  
Although there have been several theoretical calculations of the 
hadronic content of the photon \cite{do,dg,lac,grv}, at present the 
experimental data \cite{expt} are not adequate to significantly 
constrain theory.  Recently a number of papers have appeared which 
present methods of extracting the gluonic component of the photon 
\cite{eboli,bawa,gluck}.  In particular, the potential of $e\gamma$ 
colliders for the study of the hadronic content of the photon has 
been studied by several authors \cite{eboli,bawa}.  (Such colliders, 
where high energy $e^+e^-$ colliders are converted into $e\gamma$ 
colliders, are receiving considerable interest for the phenomenology 
that may be studied in $e\gamma$ collisions.)  However, these 
analysis have concentrated on dijet production in $e\gamma$ 
collisions where it is difficult to isolate the effect of the gluon 
and quark distributions.  A good determination of the quark content 
of the photon is therefore needed to test the various models of the 
photon structure functions and is essential for the extraction of 
the gluon content of the photon.  In this paper we present a novel 
method of measuring the heavy quark content of the photon using hard 
scattered $b$-quarks at $e^+e^-$ and $e\gamma$ colliders.  We find 
that by using $b$ tagging to probe quark distributions inside the 
photon it should be possible to differentiate between the various 
existing sets of photon distribution functions using a 500~GeV 
$e^+e^-$ collider operating in $e\gamma$ mode.

Recently, there has been some related work on heavy quark 
production.  In the context of leptoproduction on a proton, in 
Ref.~\cite{or} Olness and Riemersma  pointed out that there are 
advantages to using both a fixed flavor scheme (in which, 
{\it e.g.}, the $b$-quark is not considered to be a 
constituent)~\cite{lrsm} and a variable flavor scheme (in which, 
{\it e.g.}, the $b$-quark may or may not be considered a constituent 
depending on energy scale) \cite{acot}. Although these works all 
deal with leptoproduction on a proton, many of the ideas will apply, 
with suitable modifications, to leptoproduction on a photon.  In 
Ref.~\cite{htwi}, heavy quark production in two photon processes at 
an $e^+e^-$ collider is discussed; although some mention is made of 
the heavy quark content of the photon, the bulk of the analysis 
deals with the production of heavy quark pairs {\it via} 
$\gamma \gamma$, $\gamma g$ and $g g$ fusion.  Finally, Laenen and 
Riemersma~\cite{lr} present a calculation of heavy quark production 
in $e \gamma$ collisions using a fixed flavor scheme; as we consider 
the heavy quark as being a constituent of the photon, our approach 
should be considered to be complementary to that of Ref.~\cite{lr}.

\section{CALCULATION}
\label{sec:results}

We are interested in the measurement of $f_{b/\gamma}(x,Q^2)$ which 
contributes to the process $e + \gamma \to e + b + \; {\rm jet}$ via 
the subprocess $e + b \to e + b$ which can be written 
\begin{equation}
\sigma (e\gamma \to e b \; {\rm jet}) = \int dx \; 
f_{b/\gamma}(x,Q^2)
\; \hat{\sigma} (eb \to eb).
\end{equation}
This process is shown in Fig.~1.  It is straightforward to calculate 
the amplitude for the subprocess $e + b \to e + b$, and the squared 
and summed/averaged matrix element is given by:
\begin{eqnarray}
\overline{|{\cal M}|^2} = & & \nonumber \\
\nonumber \\
32 \pi^2 \alpha^2 & & \left\{ \left[(\hat{s} - m_b^2)^2 +
(\hat{u} - m_b^2)^2 \right] \left[ \frac{Q_b^2 Q_e^2}{\hat{t}^2} +
\frac{2 Q_e Q_b g_V^e g_V^b}{s_W^2 c_W^2 \hat{t}} 
    \frac{(\hat{t} - M_Z^2)}
{(\hat{t} - M_Z^2)^2 + \Gamma_Z^2 M_Z^2} \right. \right. \nonumber \\
 & + & \left. \frac{((g_V^e)^2 + (g_A^e)^2) ((g_V^b)^2 + (g_A^b)^2)}
{s_W^4 c_W^4} \frac{1}{(\hat{t} - M_Z^2)^2 + \Gamma_Z^2 M_Z^2} 
\right]
\nonumber \\
\nonumber \\
\mbox{} + 2 m_b^2 \hat{t} & & \left[ \frac{Q_b^2 Q_e^2}{\hat{t}^2} +
\frac{2 Q_e Q_b g_V^e g_V^b}{s_W^2 c_W^2 \hat{t}} 
     \frac{(\hat{t} - M_Z^2)}
{(\hat{t} - M_Z^2)^2 + \Gamma_Z^2 M_Z^2} \right. \nonumber \\
 & + & \left. \frac{((g_V^e)^2 + (g_A^e)^2) ((g_V^b)^2 - (g_A^b)^2)}
{s_W^4 c_W^4} \frac{1}{(\hat{t} - M_Z^2)^2 + \Gamma_Z^2 M_Z^2} 
\right]
\nonumber \\
\nonumber \\
\mbox{} & & \hspace*{-1.84cm} +
\left[(\hat{s} - m_b^2)^2 - (\hat{u} - m_b^2)^2 \right] \left[
\frac{2 Q_e Q_b g_A^e g_A^b}{s_W^2 c_W^2 \hat{t}} 
     \frac{(\hat{t} - M_Z^2)}
{(\hat{t} - M_Z^2)^2 + \Gamma_Z^2 M_Z^2} \right. \nonumber \\
 & + & \left. \left. \frac{4 g_V^e g_A^e g_V^b g_A^b}
{s_W^4 c_W^4} \frac{1}{(\hat{t} - M_Z^2)^2 + \Gamma_Z^2 M_Z^2} 
\right]
\right\}.
\nonumber \\
\end{eqnarray}
In the above equation, $s_W = \sin \theta_W$ and 
$c_W = \cos \theta_W$ with $\theta_W$ being the weak angle; 
$g_{V,A}^{e,b}$ are the vector and axial vector couplings of the 
electron and $b$-quark to the $Z^0$, respectively, chosen such that 
$g_A^e = 1/4$ (this fixes the remainder of the couplings 
unambiguously).  Also, $Q_{e,b}$ are the electric charges of the 
electron and $b$-quark, respectively.

In some sense the process $e \gamma \to e b \; {\rm jet}$ can be 
regarded as an approximation to the direct (pointlike) process 
$e \gamma \to e b\bar{b}$ in analogy to the effective W 
approximation or the Weisz\"acker-Williams approximation.  With 
appropriate kinematic cuts to ensure that we are describing the same 
kinematic region, the $p_T$ distributions of the outgoing $b$ in the 
two approaches closely resemble each other, as demonstrated in 
Fig.~2. In addition, contributions from the process 
$e g \to e b\bar{b}$, with the gluon coming from the hadronic 
structure of the photon, are similarly  contained in our resolved 
photon contribution {\it via} the Altarelli-Parisi evolution of the 
distribution functions.   When we consider backgrounds we must take 
care so as not to double count these contributions.

To determine whether the process is viable as a probe of the quark 
content of the photon we must include detector acceptances and 
consider the possible backgrounds.  Fortunately the signature for 
the process is quite distinct, a single $b$ balanced against the 
beam electron and possibly hadronic jet remnants of the photon. For 
$b$ detection we use a typical LEP detector acceptance for $b$'s of 
$|\cos\theta|< 0.85$ and assume a $b$ detection efficiency of 50\%.  
The jet remnants will generally go down the beam so when considering 
backgrounds we can veto events with jets detected above some minimum 
angle.  One sees that the crux of the analysis is to determine 
whether the signal is overwhelmed by backgrounds.

To illustrate the background contributions we take as an example 
$b$-quark production at a 500~GeV $e\gamma$ collider where we have 
folded in the photon spectrum for the backscattered laser and use 
(to be specific) the Duke and Owens photon structure functions 
\cite{do}.  The general properties of the other cases we will 
consider are the same.  Possible backgrounds are shown in Fig.~3.  
They can be divided into direct processes (Fig.~3a),
\begin{eqnarray}
e^+ + e^- & \to & b+ \bar{b}\\
e^+ + e^- & \to & \gamma + b + \bar{b}\\
e + \gamma & \to & e + b  + \bar{b} \\
e + \gamma & \to & \nu + b + \bar{c}
\end{eqnarray}
singly resolved (Fig.~3b)
\begin{eqnarray}
e + [g]_\gamma & \to & e + b + \bar{b} \\
e + [g]_\gamma & \to & \nu + b + \bar{c}
\end{eqnarray}
and doubly resolved (Fig.~3c)
\begin{eqnarray}
\left[g\right]_\gamma + [g]_\gamma & \to & b + \bar{b} \\
\left[q\right]_\gamma + [\bar{q} ]_\gamma & \to & b + \bar{b}
\end{eqnarray}
where we used the notation $\left[p\right]_\gamma$ to represent 
parton $p$'s content in the photon.  We include detector acceptance 
cuts of $|\cos\theta_{eb}|< 0.85$ for the $b$, where $\theta_{eb}$ 
is the angle between $b$ and the beam direction, the observed 
electron (or positron) is at least $10^o$ from the beam and all 
particles not to be detected must be within $10^o$ of the beam 
except for neutrinos.

We start with the two processes already mentioned above, 
$e \gamma \to e b\bar{b}$ (eqn. 5) and $e [g]_\gamma \to e b\bar{b}$ 
(eqn. 7).  The pointlike process $e \gamma \to e b\bar{b}$ contains 
a collinear divergence arising from the internal $b$-quark line 
which is also included in the process 
$e\gamma \to eb \; {\rm jet}$.  This divergence must therefore be 
first subtracted as described in Ref.~\cite{acot,blum} before 
including it as a background.  Likewise, the process 
$e [g]_\gamma \to e b\bar{b}$, {\it i.e.}, with the gluon coming 
from the hadronic structure of the photon, is similarly, to some 
approximation, contained in our resolved photon contribution 
{\it via} the Altarelli-Parisi evolution of the distribution 
functions.  A similar collinear subtraction must be made if 
$e [g]_\gamma \to e b\bar{b}$ is to be considered a correction to 
our result.  We find that, after making the required collinear 
subtraction, both processes $e \gamma \to e b\bar{b}$ (direct) and 
$e g \to e b\bar{b}$ (once resolved) contribute negligibly in the 
kinematic region under study.

In the remaining background discussions we are concerned with decay 
chains that might be mistaken for the signal.  In particular, in the 
processes listed in eqns. (3)-(10) heavy quarks are pair produced 
which could lead to the situation where the $\bar{b}$ (or $\bar{c}$) 
decays semileptonically and only an electron is seen by the 
detector.  Clearly, this is expected to be rather unlikely given 
that all decay products, both leptonic and hadronic, are expected to 
be boosted in the same direction given the relatively large energy 
of the produced quarks.  The likelihood of these events is further 
reduced given the small branching fractions for these decay chains.

Keeping this in mind we start by describing the possible direct 
process backgrounds.  The first two backgrounds are only relevant to 
the $e^+e^-$ cases involving Weisz\"acker-Williams photons.  The 
background $e^+ + e^- \to b + \bar{b}$ is negligible compared to 
$e^+ + e^- \to b + \bar{b} + \gamma$ even though it is technically 
of lower order in coupling.  This is because the $b$ and $\bar{b}$ 
come out back-to-back with high $p_{_T}$ and are therefore very 
distinct because the decay products of the $\bar{b}$ are boosted 
along the $\bar{b}$ direction making it very unlikely that the 
electron is seen in the detector while the hadronic remnants go down 
the beampipe.  The second direct background, 
$e^+ +e^- \to b + \bar{b} + \gamma$, will also turn out to be 
unimportant in the relevant kinematic region once the $\bar{b}$ is 
allowed to decay and cuts are imposed so that only the $e^- $ and 
$b$ are observed.  The third direct background, 
$e + \gamma \to e + b + \bar{b}$, can contribute in two ways. In the 
first, the beam electron is observed in the detector and the decay 
products of the $\bar{b}$ are not, while in the second the beam 
electron goes down the beampipe but the electron from $\bar{b}$ 
decay is seen in the detector.  As discussed above, the situation in 
which the beam electron is hard scattered into the detector is 
approximated by our signal and corrections to the signal, due to 
this process after subtracting the collinear divergence, are found 
to be negligible in this kinematic region (this process was 
calculated numerically using helicity amplitude techniques).  The 
other possibility, where the beam electron goes down the beampipe 
and is not seen in the detector, can be well described using the 
Weisz\"acker-Williams approximation and the subprocess 
$\gamma + \gamma \to b + \bar{b}$ (the subprocess cross section can 
be extracted from, {\it e.g.}, Ref. \cite{brod}).  The final direct 
process which could contaminate the signal is the charged current 
process, $e + \gamma \to \nu + b + \bar{c}$ Ref.~\cite{jikia}; here 
the electron seen in the detector comes from the decay of the 
$\bar{c}$.

We next consider the singly resolved backgrounds.  As discussed 
above, if the beam electron is hard scattered into the detector, the 
first of the singly resolved backgrounds, 
$e + [g]_\gamma \to e + b + \bar{b}$, (again, we use helicity 
amplitude techniques to calculate this process numerically), is 
approximated by our signal; corrections to our result in the 
appropriate kinematic region, after the necessary collinear 
subtraction has been made, are found to be negligible.  It is also 
possible that the beam electron goes down the beam pipe, in which 
case this background can be well described in the 
Weisz\"acker-Williams approximation with the subprocess 
$\gamma + g \to b + \bar{b}$ (this cross section can also be 
extracted from Ref. \cite{brod} by carefully modifying couplings and 
color factors).  Also, the charged current process 
$e + [g]_\gamma \to \nu + b + \bar{c}$ can also contribute to the 
background if the $\bar{c}$ decays leptonically and the resulting 
electron is observed in the detector; with some care, this 
subprocess can be extracted from Ref.~\cite{jikia}.

Finally, we consider the doubly resolved backgrounds where the 
photon from the beam electron contributes a parton from it's 
hadronic structure.  Since the parton model of photon structure is 
defined for real photons only, it is appropriate to use the 
Weisz\"acker-Williams approximation, with the subprocesses 
$[g]_\gamma + [g]_\gamma \to b + \bar{b}$ and 
$[q]_\gamma + [\bar{q}]_\gamma \to b + \bar{b}$, to calculate these 
backgrounds (see Ref. \cite{barger} for the cross sections).

The largest of the backgrounds are displayed in Fig.~4.  The 
$p_{_T}$ (of the $b$ quark) distributions of the two charged current 
processes have, as expected, rather long tails due to the large mass 
of the exchanged particle.  These $p_{_T}$ distributions are, 
however, several orders of magnitude smaller than our signal, and do 
not appear on our plot for the scale chosen.  The remaining 
subprocesses fall off rapidly with $p_{_T}$ of the $b$ quark; 
although they may produce many events at small $p_{_T}$, they are 
safely negligible for $p_{_T} > 40~GeV$.  We did not include charge 
identification of the leptons since we include both $b$ and 
$\bar{b}$ production in our distributions and estimates of event 
numbers as they both make equal contributions.  This approach also 
eliminates the complexities of $B^0 -\bar{B}^0$ mixing in the 
analysis.  

\section{RESULTS}

Having convinced the reader that the signal we are studying is 
distinct and clean of backgrounds we proceed to examine the 
sensitivity of various kinematic distributions to different 
structure function parameterizations for a number of collider 
possibilities.  We consider a group of the existing sets of photonic 
parton distributions that appear in the literature, namely the set 
of Duke and Owen (DO) \cite{do}, the set of Drees and Grassie 
(DG)~\cite{dg}, the set 1 of Abramowicz, Charchula and Levy 
(LAC)~\cite{lac} and the leading order set of Gl\"uck, Reya and Vogt 
(GRV)~\cite{grv}.

We first considered the LEP $e^+e^-$ collider at CERN upgraded in 
luminosity using the Weizs\"acker-Williams photon distributions with 
the $e\gamma$ cross sections to obtain numerical results.  The cross 
section for the signal is expected to be between 0.108~pb (GRV) to 
0.220~pb (LAC).  An order of magnitude improvement in the LEP 
luminosity, resulting in 1~fm$^{-1}$/year, would yield 108-220 $b$'s 
per year or of order 50-100 reconstructed $b$'s once efficiencies 
are included.  This small number of events is unlikely to offer an 
improvement over existing estimates of the $b$ content of the 
photon.  At the LEP200 $e^+e^-$ collider the cross section is 
expected to be 0.047~pb (GRV) to 0.092~pb (LAC) which would yield 
roughly 40 $b$'s or 20 reconstructed $b$'s for the total integrated 
luminosity of 500~pb$^{-1}$.  Clearly a higher energy and higher 
luminosity collider is needed.

%\noindent
%{\bf {$\sqrt{s}=500$~GeV$ e^+e^-$ and $e\gamma$ Colliders}}
%\subsection{$\sqrt{s}=500$~GeV$ e^+e^-$ and $e\gamma$ Colliders}

We therefore turn to the proposed higher energy NLC $e^+e^-$ 
collider in both $e^+e^-$ mode and $e\gamma$ mode with 
$\sqrt{s}$=500~GeV and very high luminosity yielding of order 
$50$~fb$^{-1}$/year.  In the $e^+e^-$ mode we fold in the 
Weizs\"acker-Williams effective photon distribution and the 
$e\gamma$ mode  we fold in the backscattered laser photon spectrum.

First, consider the $e^+e^-$ configuration.  The cross section is 
expected to be between 14~fb (GRV) and 24~fb (LAC).  Given 
50~fb$^{-1}$/yr of luminosity, we expect 700 to 1200 events per 
year, and after the 50\% $b$ reconstruction efficiency, of order 500 
reconstructed $b$'s per year.  This will allow a measurement of the 
$b$-quark distribution in the photon at some level, but even so, 
more events are desirable.  The $p_{_T}$ distribution of the 
$b$-quark is shown in Fig.~5a for various sets of photon 
distribution functions.

Lastly, consider a $\sqrt{s} = 500$~GeV linear $e^+e^-$ collider 
configured as an $e\gamma$ collider through the use of a 
backscattered laser.  In this situation, we expect a cross section 
of between 68~fb (GRV) and 138~fb (LAC), with all backgrounds 
smaller by at least  2 orders of magnitude after the implementation 
of the cuts described above.  This gives a total of 3400 to 6900 
events per year, and after the $b$ reconstruction efficiency, there 
will still be 2000-3000 events per year to analyze.  The 
$p_{_{T_b}}$ distribution for the signal is shown in Fig.~5b for 
various choices of photon distributions.

Having established that only the two cases involving an NLC can 
provide a useful measurement of the $b$-quark content of the photon 
({\it i.e.}, large enough signal combined with manageable 
backgrounds), we consider useful experimental measurements.  

Before proceeding we point out a subtlety that should be mentioned 
but that we will ignore.  The subtlety is related to the fact that 
the parton model (including the distribution function model of the 
hadronic structure of the photon) implicitly assumes that the 
constituents are massless (or, that the mass of the partons is small 
relative to all energy scales in the problem, and so can be safely 
ignored).  As we are producing a massive $b$-quark, the minimum 
$\hat{s}$ is of order $m_b^2$, and it is not necessarily possible to 
ignore the initial $b$-quark mass compared to all mass scales, 
namely $\sqrt{\hat{s}}$.  In practice it turns out that our results 
are only sensitive to a finite $m_b$ for two cases; at the lower 
energy colliders where the statistics are insufficient to make a 
meaningful measurement and for the low $x$ region of $d\sigma/dx$ 
distributions (where, following the standard parton model notation, 
$\hat{s}=xys$ and $p_b = xyp_{\bar{e}}$,  $y$ is the fraction of the 
electron's momentum carried by the photon, and $x$ is (nominally) 
the fraction of the photon's momentum carried by the $b$-quark).  In 
the latter case the uncertainties are only in the lowest $x$ region 
where the statistical errors will still overwhelm the uncertainties 
in our definition of the scale.

In the figures that follow, only the estimated {\bf statistical} 
uncertainties are included in the error bars, so that the error bars 
shown will slightly underestimate the actual errors.  

A physically measurable quantity is the ratio of the initial 
$b$-quark energy or momentum to the beam energy.  This will 
generally be closely related to $\tau = x y$, and the conversion is 
easily calculated.  We show, in Fig.~6a, the number of events 
{\it vs.} $\tau$ for various photon distribution functions in the 
$e\gamma$ case.  The largest differences in the models are in the 
lowest $\tau$ bins where the event numbers are largest.  Considering 
only the lowest $\tau$ bin and using the Duke-Owens results to 
estimate the error we obtain a statistical error of roughly 2.5\%.  
Thus, using  only the lowest $\tau$ bin for the measurement, and the 
higher $\tau$ bins as a normalization, it will be quite easy to 
distinguish between the various distribution functions.  Fig.~6b 
shows the same distribution in the $e^+e^-$ case.  Here, the 
conclusion is not so clear, due to relatively larger error bars, but 
it should be possible to distinguish between LAC, DO/DG and GRV, 
though it may not be possible to distinguish DO from DG.  

Finally we show, in Fig.~6a ($e\gamma$) and in Fig.~6b ($e^+e^-$) 
the distribution of event numbers in $x$.  This distribution will be 
useful if one can tag the electron which provides the 
Weizs\"acker-Williams photon (allowing a complete reconstruction of 
the $e\gamma$ initial state, including $y$), or in the $e\gamma$ 
case if one can deconvolute the $\tau$ distribution knowing the 
parton level cross section and the laser backscattered photon 
spectrum.  As with the $\tau$ distribution, the large $x$ bins can 
be used to fix the normalization of the distribution while the 
lowest $x$ bin can be used to provide information on the $b$-quark 
distribution in the photon.  Similar conclusions can be reached here 
as well: in the $e\gamma$ case, it should be possible to distinguish 
between the various distribution functions, while in the $e^+ e^-$ 
case it should be possible to distinguish between LAC, DO/DG and GRV 
(though possibly not between DO and DG).

\section{CONCLUSIONS}
\label{sec:conclusions}

In this paper we proposed  using high $p_{_T}$ $b$-quarks as a means 
of determining the heavy quark content of the photon.  We have shown 
that the process $e+\gamma \to e + b + X$ provides a clean method of 
extracting the quark content of the photon and can distinguish 
between the different models in the literature.  The subprocess we 
examined was $e+b\to e+b$ where the beam electron undergoes a hard 
scatter from a resolved $b$ in the photon, in essence Rutherford 
scattering off a photon target.  The signal for the process is quite 
distinct; all backgrounds can be eliminated by insisting that only 
the scattered $b$-quark balanced against the beam electron be 
observed and imposing a minimum $p_{_T}$ cut on the $b$ to eliminate 
the remaining direct and once resolved $b\bar{b}$ production.

It is clear that this reaction will allow a good determination of 
the quark content of the photon.  In the low $x$-regions of the 
distributions, where the models differ the most, typical 
uncertainties are roughly 2.5\% for the $e\gamma$ colliders.  This 
is more than adequate to distinguish between different models.  For 
the $e^+e^-$ cases the statistics are poorer and do not offer quite 
as good a measurement, though valuable information can be extracted 
from these data.  One could also use this approach to study the 
charmed quark content of the photon~\cite{miller}.  The $c$-quark 
content of the photon is roughly four times as large as the 
$b$-quark content due to the different quark electric charges.  
However, since charmed mesons are more difficult to reconstruct, the 
$b$-quark gives a cleaner signal.  The process we have proposed is 
complementary to others that have been studied in the literature and 
can contribute to more precise determinations of quark and gluon 
distributions inside the photon.  Along with other processes 
considered elsewhere it is clear that an $e\gamma$ collider is an 
ideal facility for studying the hadronic content of the photon.

\acknowledgments

SG thanks Rohini Godbole for helpful discussions in the early stages 
of this work and Richard Hemingway for helpful conversations.  MAD 
thanks Manuel Drees for many enlightening discussions during the 
course of this work.  This research was supported in part by the 
Natural Sciences and Engineering Research Council of Canada (NSERC), 
and the work of MAD was supported in part by an NSERC Canada 
International Fellowship.

\figure{Feynman diagram for the process $e\gamma \to eb+X$.}

\figure{Comparison of direct $e + \gamma \to e + b + \bar{b}$ 
(dashed curve), $e + [g]_\gamma \to e + b + \bar{b}$ (dotdashed 
curve) and the sum (dotted curve) to the signal process 
$e + \gamma \to e + b + X$ (solid curve).  All curves correspond to 
a 500~GeV $e^+e^-$ collider operating in $e \gamma$ mode.  For the 
signal, $e[b]_\gamma \to eb$, (solid curve) we require that the 
final state electron and $b$-quark be seen, $\theta_e$ more than 
$10^o$ from the beampipe and $|\cos \theta_b | < 0.85$; for the 
direct processes, there is an additional requirement, 
$| \cos \theta_{\bar{b}} | > 0.85$.}

\figure{Diagrams for background processes. (a) Direct contributions; 
(b) Singly resolved contributions; (c) Doubly resolved 
contributions.}

\figure{$p_{_{T_b}}$ distribution for the signal (solid line) and 
the most dangerous backgrounds: 
$\gamma + [g]_\gamma \to b + \bar{b}$ (rightmost dotted line), 
$\gamma + \gamma \to b + \bar{b}$ (rightmost dashed line), 
$[g]_\gamma + [g]_\gamma \to b + \bar{b}$ (leftmost dotted line), 
$[q]_\gamma + [\bar{q}]_\gamma \to b + \bar{b}$ (leftmost dashed 
line), $e^+ + e^- \to \gamma + b + \bar{b}$ (dotdashed line).  These 
are all make use of the Duke and Owens photon distribution functions 
(where appropriate), and are for a 500~GeV $e^+ e^-$ collider 
operating in $e\gamma$ mode.}

\figure{$p_{_{T_b}}$ distribution for the signal with LAC (solid 
line), GRV (dashed line), DO (dotted line) and DG (dotdashed line) 
distributions, using $Q^2 = \hat{s}$ at (a) a $500~GeV$ $e^+ e^-$ 
collider and (b) a $500~GeV$ $e^+ e^-$ collider configured as an 
$e \gamma$ collider.}

\figure{Event number {\it vs.} $\tau$ for a 500~GeV $e^+ e^-$ 
collider assuming L=50~fb$^{-1}$ operating in (a) $e\gamma$ mode and 
(b) $e^+ e^-$ mode, using the four different sets of photon 
distribution functions.  The solid curve is for LAC set 1, the 
dashed curve for GRV, the dotted curve for DO and the dotdashed 
curve for DG.}

\figure{Event number {\it vs.} $x$ for a 500~GeV $e^+ e^-$ collider 
assuming L=50~fb$^{-1}$ operating in (a) $e\gamma$ mode and (b) 
$e^+ e^-$ mode, using the four different sets of photon distribution 
functions.  The solid curve is for LAC set 1, the dashed curve for 
GRV, the dotted curve for DO and the dotdashed curve for DG.}

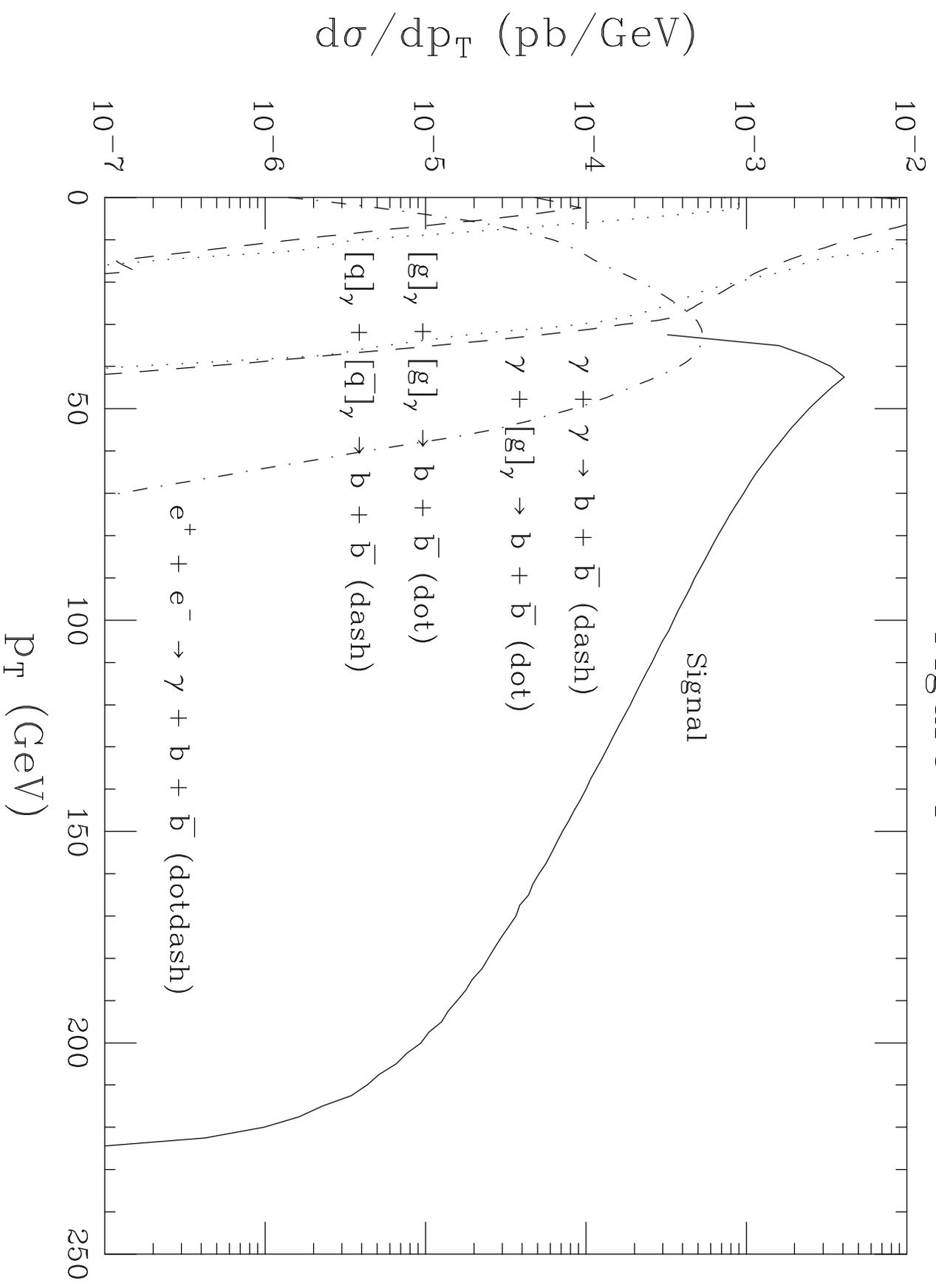
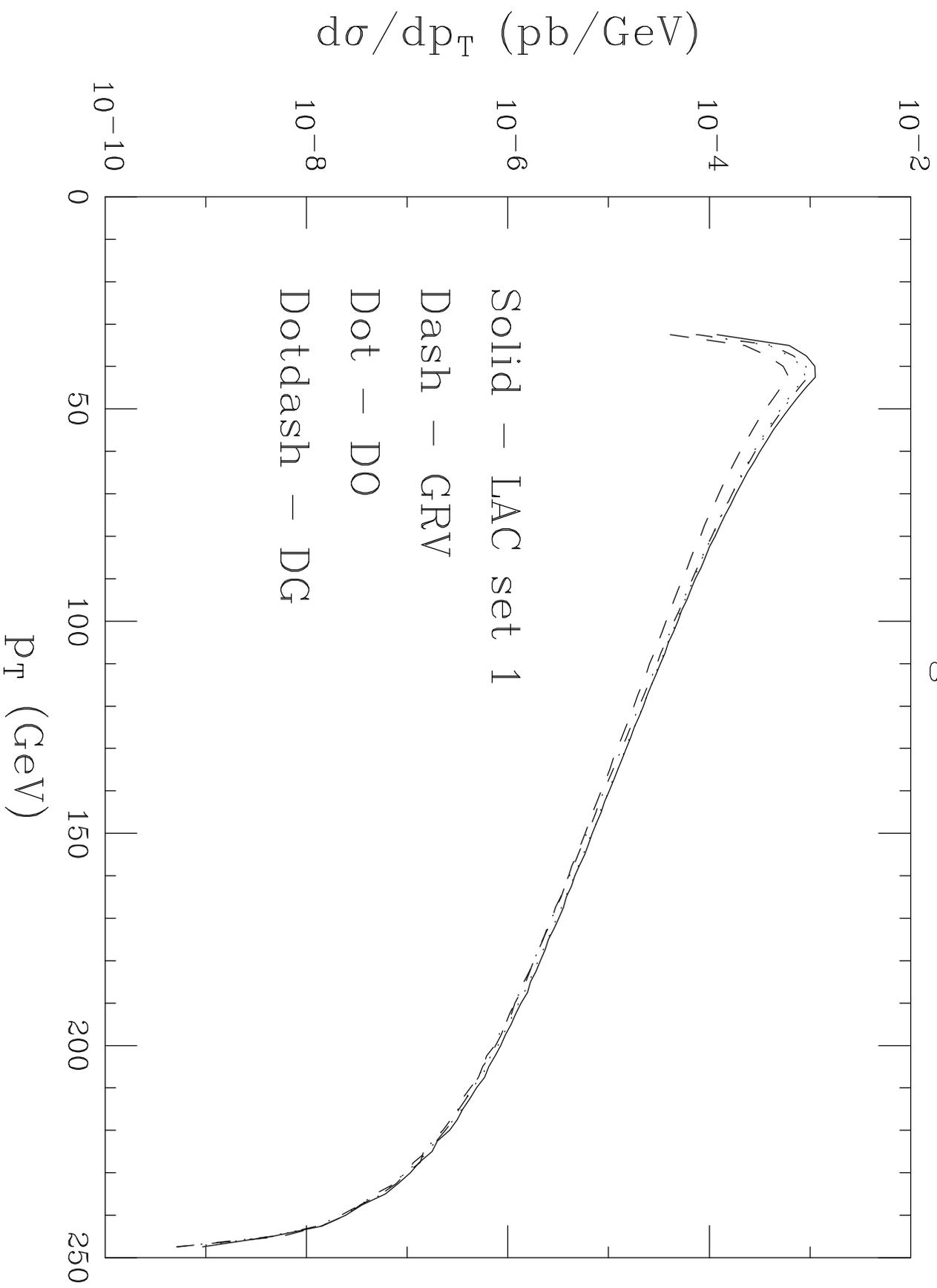
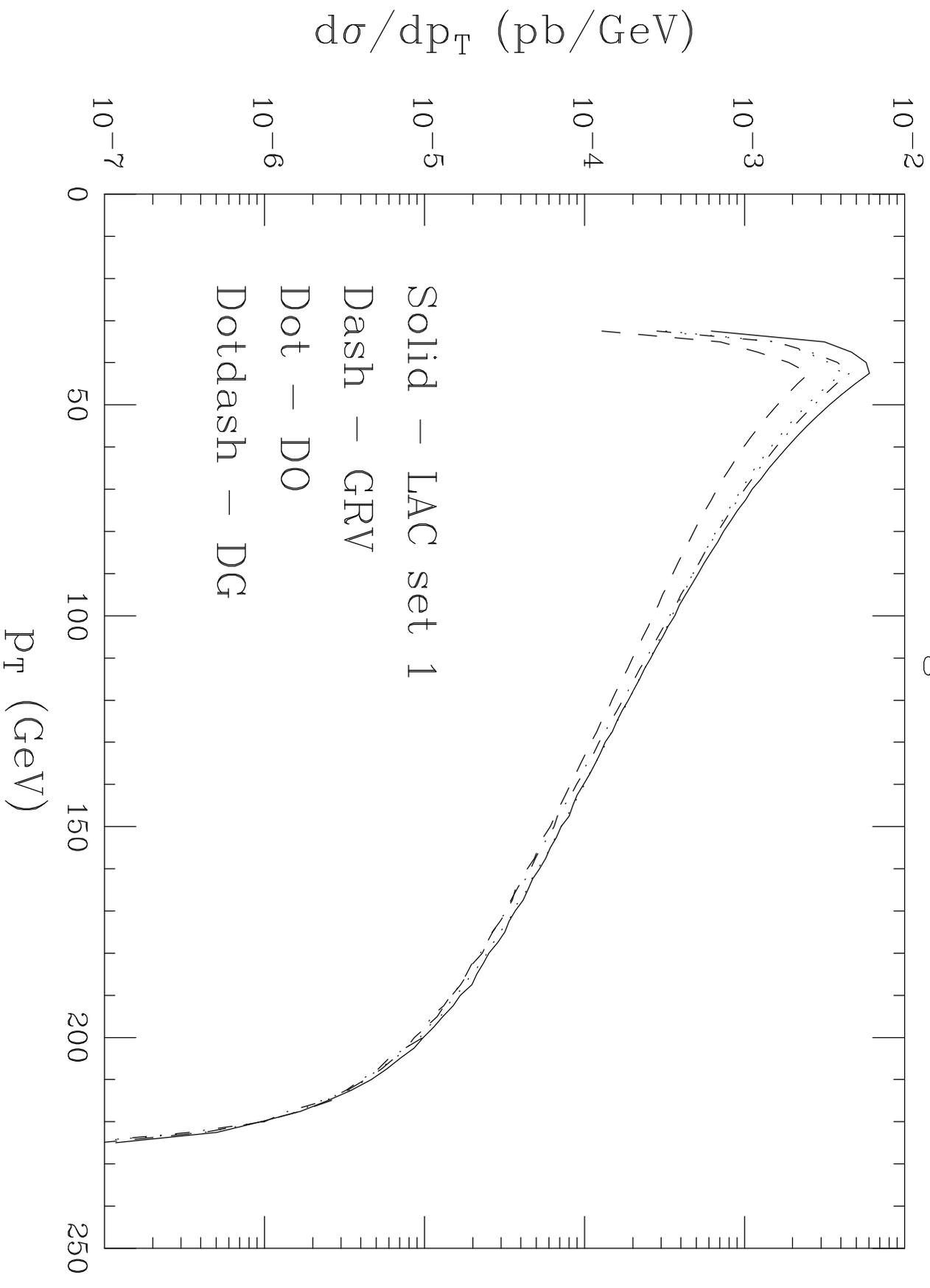
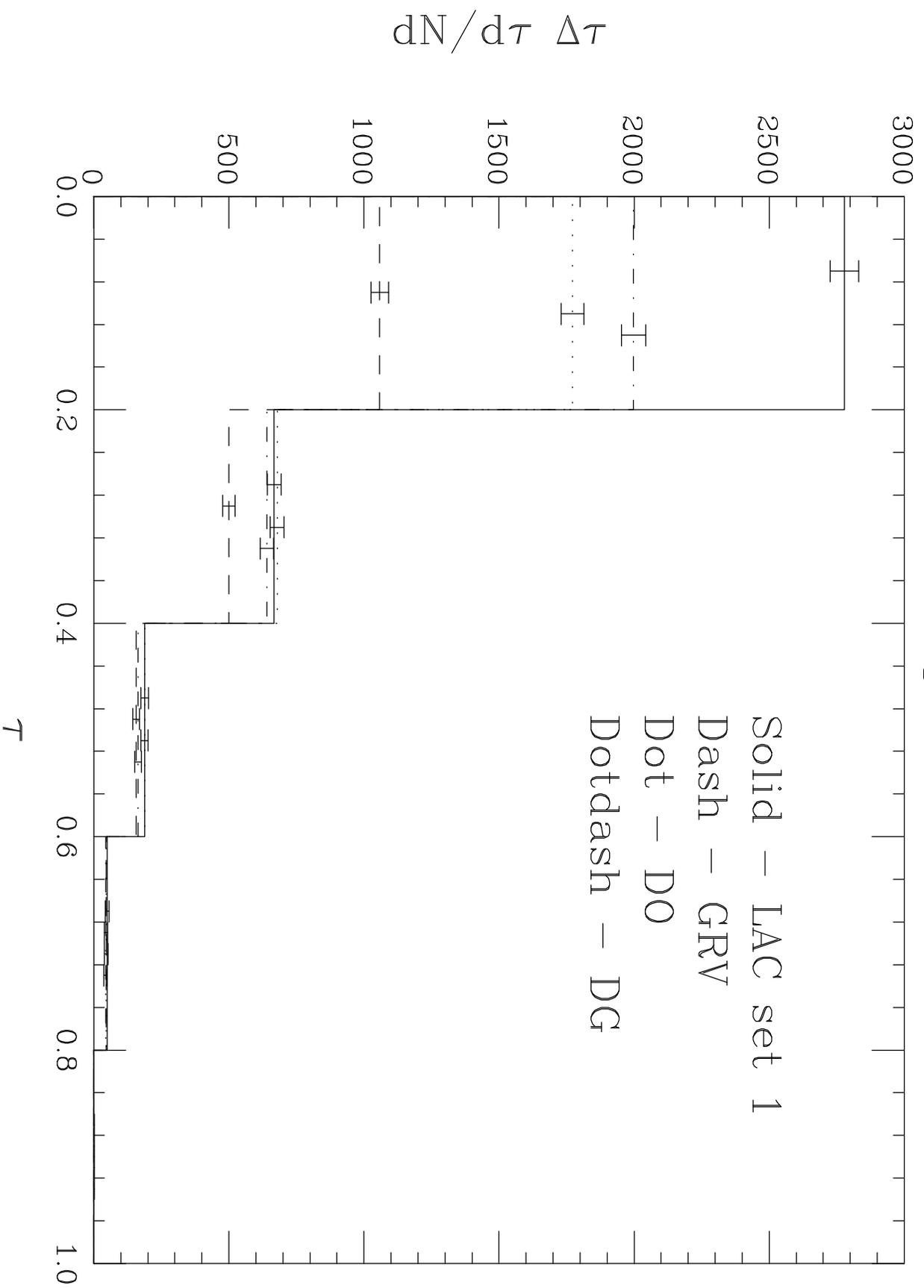
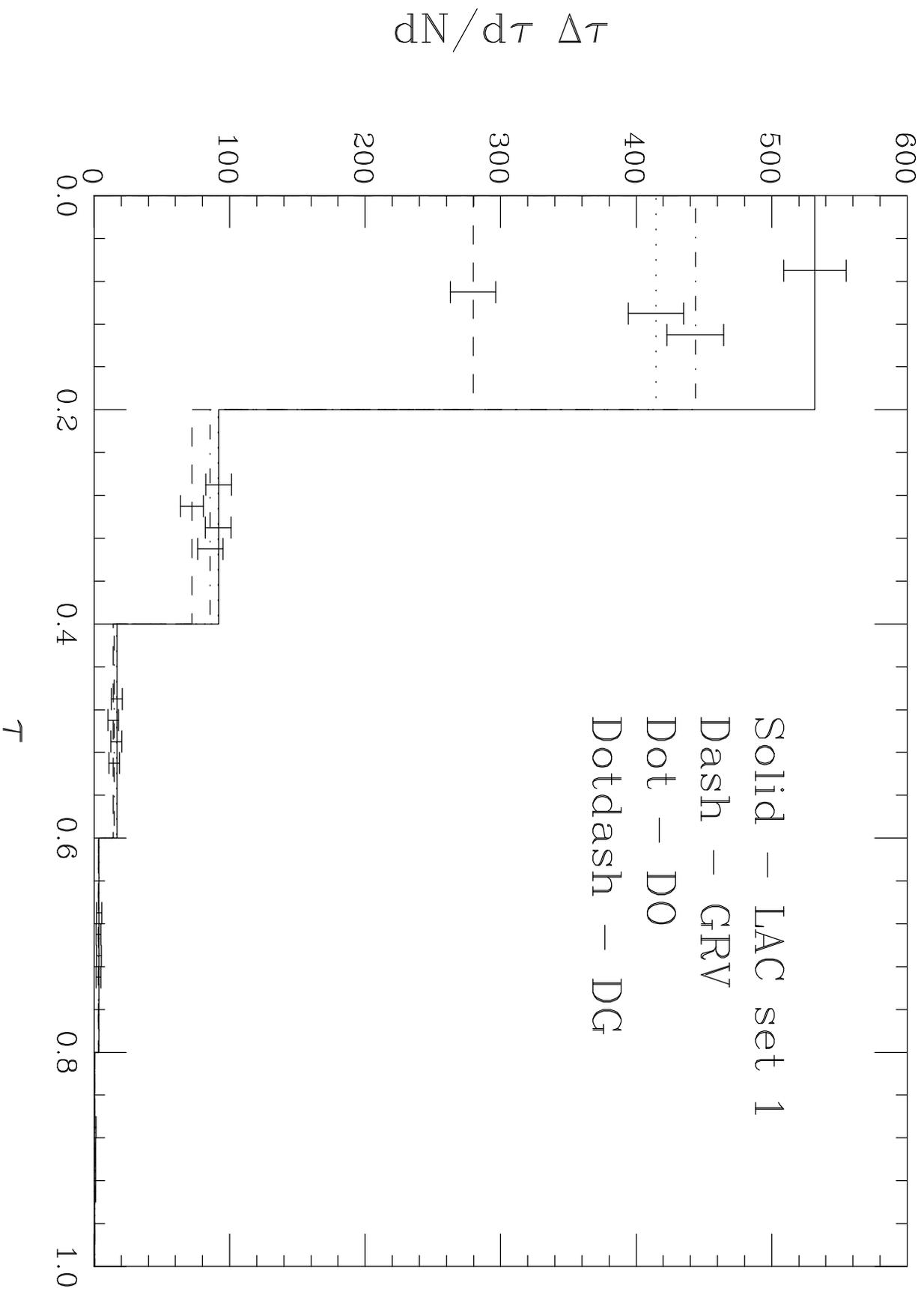
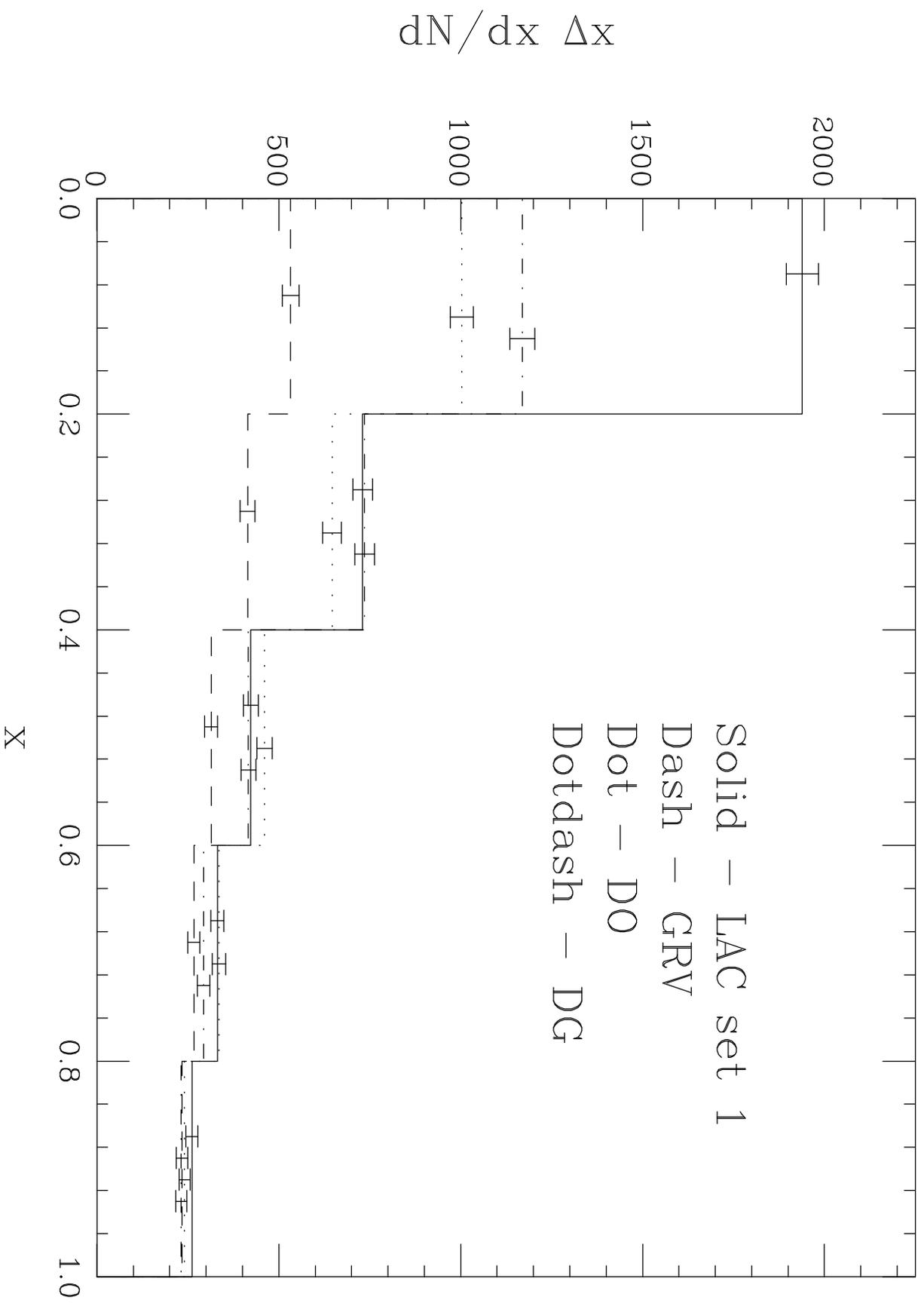
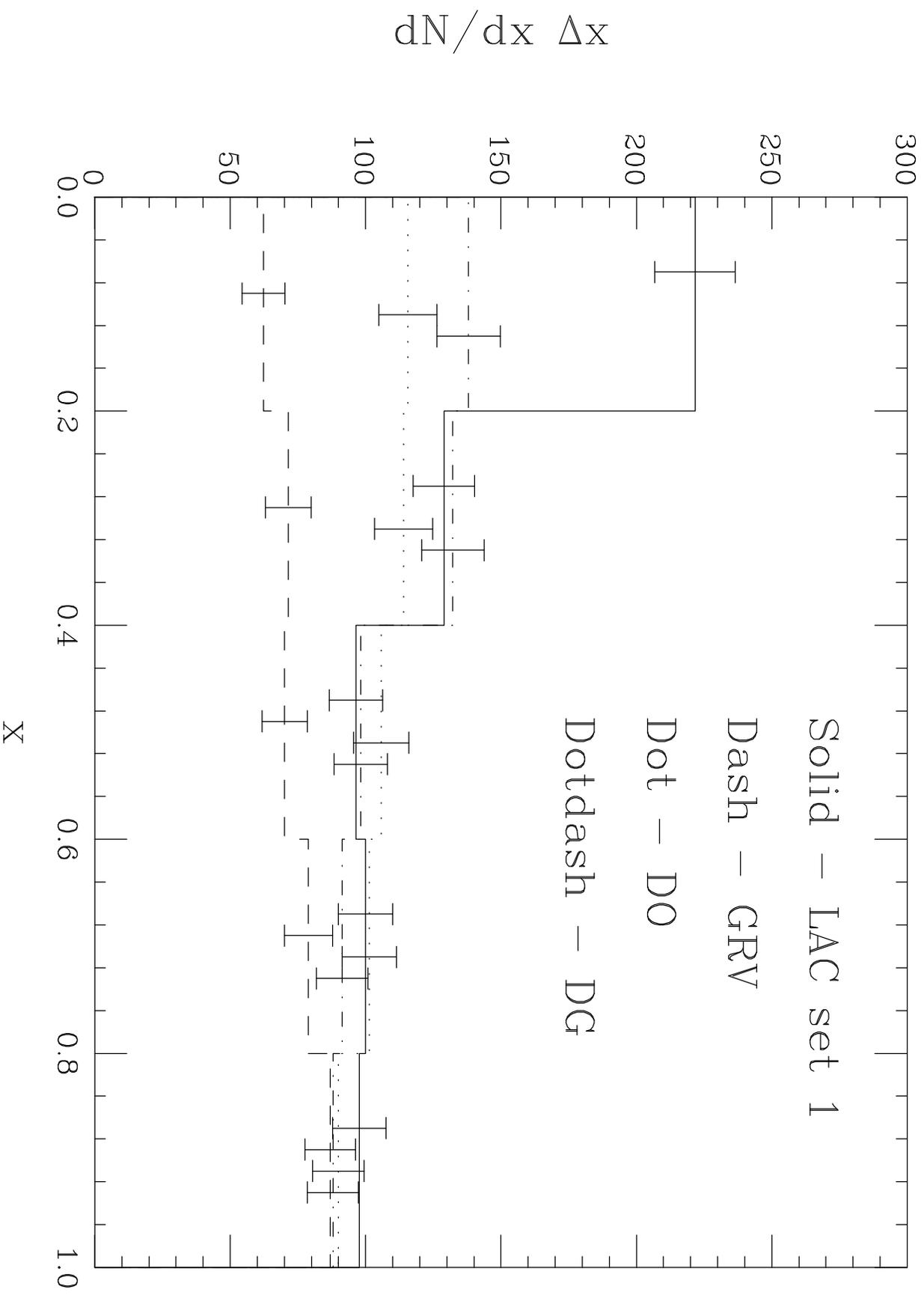
\end{document}